\documentclass[a4paper,twocolumn,11pt,accepted=2025-10-14]{quantumarticle}
\pdfoutput=1
\usepackage[utf8]{inputenc}
\usepackage[english]{babel}
\usepackage[T1]{fontenc}
\usepackage{amsmath}
\usepackage{hyperref}

\usepackage{tikz}
\usepackage{lipsum}
\usepackage{amsmath,amssymb}
\usepackage{qcircuit}

\newcommand{\Tr}{{\rm Tr}}
\def\ket#1{ | #1 \rangle}

\begin{document}

\title{Efficient Classical Simulation of the DQC1 Circuit with Zero Discord}

\author{Shalin Jose}
\affiliation{School of Physics, Indian Institute of Science Education and Research Thiruvananthapuram, Maruthamala PO, Vithura, Kerala, 695551, India}
\email{shaliniiser16@iisertvm.ac.in}
\orcid{0000-0002-2371-4327}

\author{Akshay Kannan Sairam}
\affiliation{Department of Instrumentation and Applied Physics, Indian Institute of Science, Bengaluru, Karnataka, 560012, India
} 
\orcid{0000-0003-0194-409X}

\author{Anil Shaji}

\affiliation{School of Physics, Indian Institute of Science Education and Research Thiruvananthapuram, Maruthamala PO, Vithura, Kerala, 695551, India} 
\affiliation{Center for High Performance Computing, Indian Institute of Science Education and Research Thiruvananthapuram, Maruthamala PO, Vithura, Kerala, 695551, India
} 
\orcid{0000-0002-1574-8885}
\maketitle

\begin{abstract}
  A path for efficient classical simulation of the DQC1 circuit that estimates the trace of an implementable unitary under the zero discord condition [Phys.~Rev.~Lett.~\textbf{105}, 190502 (2010)] is presented.  This result reinforces the status of non-classical correlations quantified by quantum discord and related measures as the key resource enabling exponential speedups in mixed state quantum computation. 
\end{abstract}

\section{Introduction}
Identifying the resources that allow quantum algorithms to perform certain tasks significantly faster than the best known classical ones is an increasingly important question now that small scale quantum computers are already in operation \textcolor{black}{~\cite{Arute2019, Castelvecchi2023}}. The framework of quantum information theory remains incomplete without pinpointing the quantum resources responsible for this computational advantage in each case. Within the paradigm of pure state quantum computation, multipartite entanglement that grows unboundedly with the input size was proven to be a necessary resource~\cite{jozsa2003role}. On the other hand, the Gottesman-Knill theorem \cite{aaronson2004improved} showed that highly entangled stabilizer states can be simulated efficiently using classical means and therefore cannot be a resource for quantum speedup. Just the presence of large quantities of entanglement is therefore not a sufficient condition for computational speedup and quite a bit of exploration has been done on whether particular quantum circuits can be simulated classically~\cite{vidal2003efficient, Bravyi2019simulationofquantum, hakop2015sampling, Jackson2022partition}.

Certain models of mixed state quantum computation, conceived in the context of NMR based experiments~\cite{Braun1999NMR}, like the DQC1 model~\cite{knill1998Dqc1}, can perform specific computations exponentially faster than classical analogues. Since little, or no, entanglement is generated in the state of the quantum computer during the computation in such models, entanglement alone cannot explain the speedup~\cite{Meyer2000QS}. The key resource enabling the quantum advantage in mixed state quantum computation remains an open question, with the variety of quantum phenomena proposed as candidate resources growing with time \cite{bermejo2017contextuality, veitch2012negative, jozsa2003role}. A promising candidate is quantum discord~\cite{Zurek2001discord, henderson2001classical}, which will be our focus here. Demonstrating the presence of discord in the DQC1 model was one major step in establishing it's relevance as the key quantum resource~\cite{datta2008quantum}. However, Dakic et al. \cite{dakic2010necessary} conjectured that DQC1 may provide an exponential speed-up even when discord is zero, suggesting that it may not be the pertinent resource. While this conjecture has received some attention in the past~\cite{datta2011quantum, 2016Gu, eastin2010} no conclusive statement about it has been made.

In this paper, we show that the zero discord DQC1 computation discussed in~\cite{dakic2010necessary} can be simulated efficiently by classical means \textcolor{black}{provided the DQC1 computation itself is feasible in the sense that an efficient, gate-based implementation of the corresponding circuit is available.} Central to the DQC1 model is a large unitary matrix acting on $n$ qubits which we are assuming can be decomposed into elementary gates. The purpose of the computation is to find the trace of this unitary. To show that an efficient classical simulation of the computation is possible, we first derive a relation between the trace of the zero-discord case unitary and it's diagonal elements implying that measurement of certain matrix elements contain information about the trace. Adding to this, we prove that, with a tractable principle branch Hamiltonian for the corresponding unitary, there is an efficient classical algorithm for finding these matrix elements leading to the trace. Our results restores the status of quantum discord as a viable quantifier of the key resource that enables exponential speedup in mixed state quantum computation.

\section{DQC1 and Quantum Discord}
\begin{figure}[h!]
	\includegraphics[scale = 0.5]{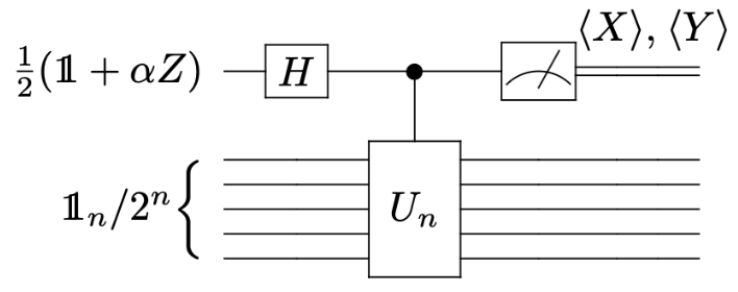}
	\caption{The DQC1 circuit for evaluating the normalized trace of the unitary $U_n$ acting on the bottom register of $n$-qubits which are initialized in the fully mixed state.  } 
	\label{fig:dqc1}
\end{figure}

The DQC1 circuit shown in fig (\ref{fig:dqc1}) consists of a single qubit of non-zero purity in the state $(\openone + \alpha Z)/2 $, coupled to an $n$-qubit register in completely mixed state $\openone_{n}/2^{n}$. In the following the three Pauli spin operators are denoted by $X$, $Y$ and $Z$ respectively. It can efficiently evaluate the normalized trace of the unitary matrix $U_n$ that acts on the $n$-qubit fully mixed state. As mentioned above, it is assumed that $U_n$ is implementable in poly($n$) time. This model was first introduced by Knill and Laflamme \cite{knill1998Dqc1} and has been shown to provide exponential speed up for other related problems as well~\cite{poulin2004fidelity,knill2001quantum,Shor2008Jones,boixo2008parameter}.
The output state of the circuit is given as,
\begin{equation*}
    \label{DQC1state}
    \rho = \frac{1}{2^{n+1}} \bigg( \openone_1 \otimes \openone_n + \alpha|0\rangle \langle 1|\otimes U_n^\dagger + \alpha|1\rangle \langle 0|\otimes U_n \bigg)
\end{equation*}

The normalized trace of $U_{n}$, $\tau = \Tr(U_{n})/2^{n}$, can be estimated by performing measurements of $X$ and $Y$ operators on the first qubit with $\tau_{r} = \langle \sigma_x \rangle / \alpha = \text{Re}(\Tr(U_{n})/2^{n}) $ and $\tau_{i} =  \langle \sigma_y \rangle / \alpha = \text{Im}(\Tr(U_{n})/2^{n}) $. To estimate the trace to any fixed, desirable precision, a fixed number of repetitions of the measurements on top qubit are sufficient. This is in contrast to the resource requirement for any known classical method for evaluating the trace of the $2^n \times 2^n$ unitary $U_n$, all of which scale exponentially with $n$. The DQC1 circuit therefore achieves an exponential speedup but bipartite entanglement across the natural division between the top qubit and the rest is never generated during any step of the DQC1 algorithm. Across other bi-partitions some minimal amount of entanglement is present but it is bounded and is vanishingly small in the limit of $n\rightarrow \infty$~\cite{datta2005EntanglementDQC1}.  However finite amounts of non-classical correlations as quantified by quantum discord were observed to be present and it was proposed that such correlations may be the key quantum resource enabling the speed up~\cite{datta2008quantum}.
\subsection{Quantum Discord}
For a composite quantum system $AB$ with constituent sub-systems $A$ and $B$ described by the density matrices $\rho_{AB}$, $\rho_A$ and $\rho_B$ respectively, Quantum discord, ${\mathcal D}(B,A)$ captures the quantum correlations between $A$ and $B$ as the difference between total correlations and the purely classical correlations. Total correlations are quantified using the quantum mutual information given as, 
\begin{equation}
    {\mathcal I}(B:A) = H(\rho_B) + H(\rho_A) - H(\rho_{AB})
    \label{eq:mutual}
\end{equation}
where $H(\rho) = - {\rm tr}(\rho \log \rho)$ is the von Neumann entropy of $\rho$. Classical correlations between the two subsystems are captured by,
\begin{equation}
    \mathcal{J}(B:A) = H(B) - \Tilde{H}(B|A) 
    \label{Jclassical}
\end{equation}
where  $\tilde{H}(B|A) = \min_{\{\Pi_i^A\}} \sum_i p_i  H(\rho_{B|i})$ with the minimisation of the conditional entropy $H(S|M) = \sum_i p_i  H(\rho_{S|i})$ taken over all possible measurements on $A$. In practice, the measurements are limited to projective ones for computational simplicity. This leads to quantum discord given as, 
\begin{equation*}
    \mathcal{D}(B,A) = \mathcal{I}(B:A) - \mathcal{J}(B:A)
\end{equation*}
We are specifically interested in the case where the state of the $n+1$ qubits in the DQC1 circuit after the application of $U_n$ has zero discord across top qubit versus rest division with measurements on the top qubit. In~\cite{dakic2010necessary} it was shown that given a singular value decomposition of a state $\rho_{AB}$ as $\sum_k c_k A_k \otimes B_k$ with $A_k$ and $B_k$ being operators belonging to the Hilbert spaces of $A$ and $B$ respectively, ${\mathcal D}(B:A)$ is zero if there exists a set of projection operators $\{\Pi_j^A\}$ such that $\sum_j\Pi_j^A A_k \Pi_j^A = A_k$. This also implies $[A_j,A_k]=0$. The DQC1 state after $U_n$ can be written in the form,
\begin{align}
\rho_{AB} = \frac{1}{2^{n+1}} \bigg( \openone_1 \otimes \openone_n + \alpha X \otimes \frac{U_n + U_n^\dagger}{2} \nonumber \\ 
+ \alpha Y \otimes  \frac{U_n - U_n^\dagger}{2i} \bigg), 
\end{align}
where $X$, $Y$ and $Z$ are the three Pauli spin operators. This means that the DQC1 state after $U_n$ is a zero discord state provided $U_n^\dagger = kU_n$ which is possible only if $U_n = e^{i\phi} O_n$ with $O_n^2 = \openone$. In~\cite{dakic2010necessary} it is conjectured that finding the trace of $U_n = e^{i\phi} O_n$ efficiently using classical means may not be possible for all such unitaries while DQC1 can still perform the task efficiently thereby implying that  discord (which is zero) cannot be a quantifier for the key resource behind exponential quantum speedup. 

The question of classical simulation of the zero discord DQC1 model is now reduced to trace estimation of the restricted class of unitary matrices with $U_n^\dagger = kU_n$. We show in the following that $U_n$ is sufficiently restricted with $O_n$  being involutory, that it is possible to estimate its trace classically in an efficient manner as long as $U_n$ is also assumed to be efficiently implementable as it should be for any realistic DQC1 circuit.

\section{Trace Estimation}
Estimation of the global phase of $U$, if any, will be discussed separately. First we focus on finding the magnitude of the trace of $U$ setting $\phi = 0$. Note that we have dropped the subscript $n$ from the operators $U$ and $O$ in the following for simplifying the notation. The Hermitian Unitary matrix $O$ has $\det(O) = \pm 1$ and eigenvalues $\lambda_j = \pm 1$ so that,
\begin{equation}
    {\rm Tr[O]} = N_{+} - N_{-},
\label{eq:trace}
\end{equation}
where $N_{+}$ and $N_{-}$ are the number of eigenvalues of $O$ equal to $+1$ and $-1$ respectively. \textcolor{black}{The Hermitian unitary $O$ can always be written as} $O = V^{\dagger} D V$, where $D$ is a diagonal matrix with $+1$ and $-1$ as entries while $V$ is a Haar-random unitary. Since the Haar measure is invariant under permutation of rows and columns~\cite[p.~138]{hiai2000semicircle}, we assume that $D$ has been rearranged through a suitable permutation of rows and columns with all the $-1$ entries clubbed together at the top so that we can write it in the form $D={\rm diag}[-1,-1,..,1,1]$. With this rearrangement, the $q^{\rm th}$ diagonal element of $O$ is, 
\begin{equation}
    \label{eq:oqq}
    O_{qq} = \sum_{k}|v_{qk}|^{2}d_{k} = \sum_{k > n_{-1}}|v_{qk}|^{2} - \sum_{k \leq n_{-1}}|v_{qk}|^{2}, 
\end{equation}
where $d_{k} \in [1,-1]$ denote the diagonal entries of $D$ and $v_{qk} = \langle q|V|k\rangle$ with $| q \rangle$ and $|k\rangle$ being the (permuted) computational basis vectors. 
\begin{figure}[!htb]
    \centering
    \includegraphics[width=0.8\linewidth]{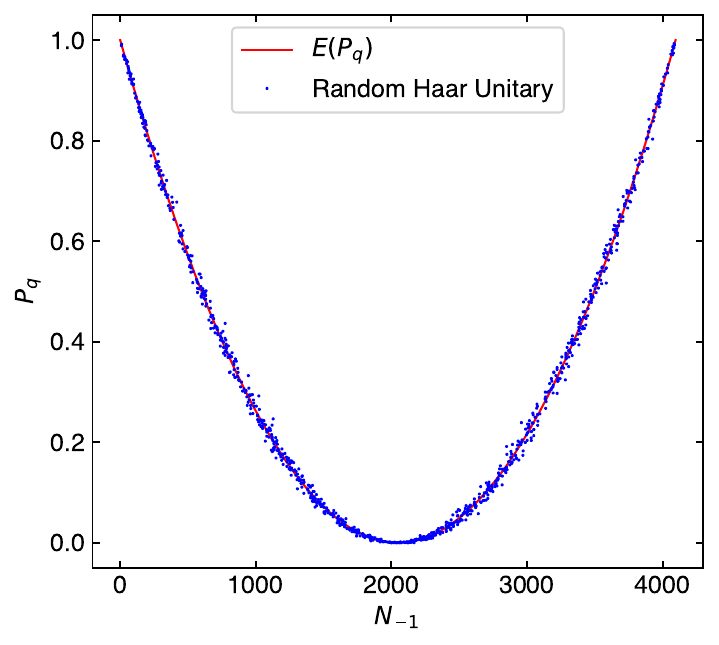}
\caption{Probability, $P_q$ of measuring $\ket{00\cdots00}$ at the output of a randomly generated 12-qubit zero discord unitary, when $\ket{00\cdots0}$ is the input state is plotted against $N_-$. Also shown as the solid line (red) is the analytic expression for the expectation value of $P_q$ from Eq.~\eqref{eq:var}. We see that the numerically obtained points are distributed around the expected value with small variance.}
\label{fig:P_state}
\end{figure}
We now consider the expectation value of $O_{qq}$ with respect to the Haar probability measure $\mu_N$ over the unitary group $\mathcal{U}_{N}$ with $N=2^n$. For any measurable function $f$ over $\mathcal{U}_N$, the expectation value is defined as ${\mathbb E}(f) = \int_{V \in \mathcal{V}_{N}}f(V) d\mu_{N}(V)$. The following results from Proposition 4.2.3 of~\cite{hiai2000semicircle} applies to any Haar-random unitary:
\begin{eqnarray} 
\label{eq:petz}
{\mathbb E}(|v_{ij}|^2) & = &  \frac{1}{N},  \nonumber \\
{\mathbb E}(|v_{ij}|^4) & = & \frac{2}{N(N+1)}, \nonumber \\
{\mathbb E}(|v_{ij}|^2|v_{ij'}|^2) & = & \frac{1}{N(N+1)}, \;\; (j \neq j').
\end{eqnarray}
Using the equations above, we obtain,
\begin{equation}
    \label{eq:means1}
    {\mathbb E}(O_{qq}) = \frac{N_+ - N_-}{N},
\end{equation}
which is nothing but the normalised trace of $O$. If any computational basis state $|q\rangle$ is inserted as the input into the bottom half of the zero-discord DQC1 circuit, then the probability of obtaining the same state as output is given by, 
\begin{align}
    \label{Pn}
        P_{q}  = & |\langle q|U_N | q \rangle|^{2} = O_{qq}^2 \nonumber \\
         = & \Big(\sum_{k}|u_{qk}|^{2}d_{k}\Big) \Big(\sum_{l}|u_{ql}|^{2}d_{l}\Big) \nonumber \\
         = & \sum_{k,l\leq N_-} \!\! |v_{qk}|^{2} |v_{ql}|^{2} - \!\!\!\!\! \sum_{k> N_-,l\leq N_-} \!\!\!\!\!\!\! |v_{qk}|^{2}  |v_{ql}|^{2} \nonumber \\
         & \quad  - \!\!\!\!\!\! \sum_{l> N_-,k\leq N_-}  \!\!\!\!\!\!\! |v_{qk}|^{2} |v_{ql}|^{2}
        + \sum_{k,l> N_-} \!\! |v_{qk}|^{2} |v_{ql}|^{2}
\end{align}
Using Eq.~\eqref{eq:petz} we obtain, 
\begin{equation}
    \label{eq:var}
    {\mathbb E}(P_q)  =  {\mathbb E}(O_{qq}^2) = 
     \frac{1}{N+1} + \frac{(N_+ - N_-)^2}{N(N+1)}.   
\end{equation}
From Eqs.~\eqref{eq:oqq}, \eqref{eq:means1}, \eqref{Pn} and \eqref{eq:var} we observe the following about the zero discord case. The magnitude, apart from an overall phase, of the normalized trace of $U_N$ can be obtained from ${\mathbb E}(O_{qq})$  only and $O_{qq}^2 = P_q$ is an observable quantity. Furthermore, with respect to the Haar measure, $O_{qq}$ is a random variable distributed with variance,  $\Delta(O_{qq}) = {\mathbb E}(O_{qq}^2) - [{\mathbb E}(O_{qq})]^2 \sim 1/N$, for large $N$. The means that when the number of qubits in the bottom register of DQC1 circuit is moderately big so that $N = 2^n$ is large, the estimate of $O_{qq}$ obtained from $P_q$ is equal to ${\mathcal E}(O_{qq})$ with very high confidence level. We can see this from Fig.~\ref{fig:P_state}, where the probability of obtaining the state $|00\cdots 00\rangle$ at the output end of the DQC1 circuit, corresponding to the same state as input, is plotted as a function of $N_-$ for randomly generated 8-qubit ($N=256$) zero discord unitaries. Also shown in the figure is ${\mathbb E}(P_{00\cdots 00})$ from Eq.~\eqref{eq:var} written as a function of only $N_-$ using $N_+ = N-N_-$.

\textcolor{black}{
In the discussion above we require the unitary $V$ in $V^{\dagger} DV$ to be Haar random for Eq.~\eqref{eq:var} to hold and for $P_q$ to converge to a parabola with respect $N_{-1}$ as in Fig.~\ref{fig:P_state}. This raises the following question. What if the the DQC1 unitary is highly structured and from inspection itself it is suspected that $V$ is not a typical Haar random unitary. This question can be addressed by noting that changing $V$ does not change the eigenvalues of $O$ or the trace of $O$. Furthermore the $V^{\dagger} DV$ form allows us to pad the given implementation of the unitary with layers of gates on either side with corresponding layers being adjoint of each other without changing the $V^{\dagger} DV$ form. By randomly choosing the gates in these padding layers we can make the unitary sandwiching $D$ effectively Haar-random. Since we are adding only finite number of layers, the padding does not increase the computational complexity of our classical simulation. }
\begin{figure}[!htb]
    \centering
    \includegraphics[width=1\linewidth]{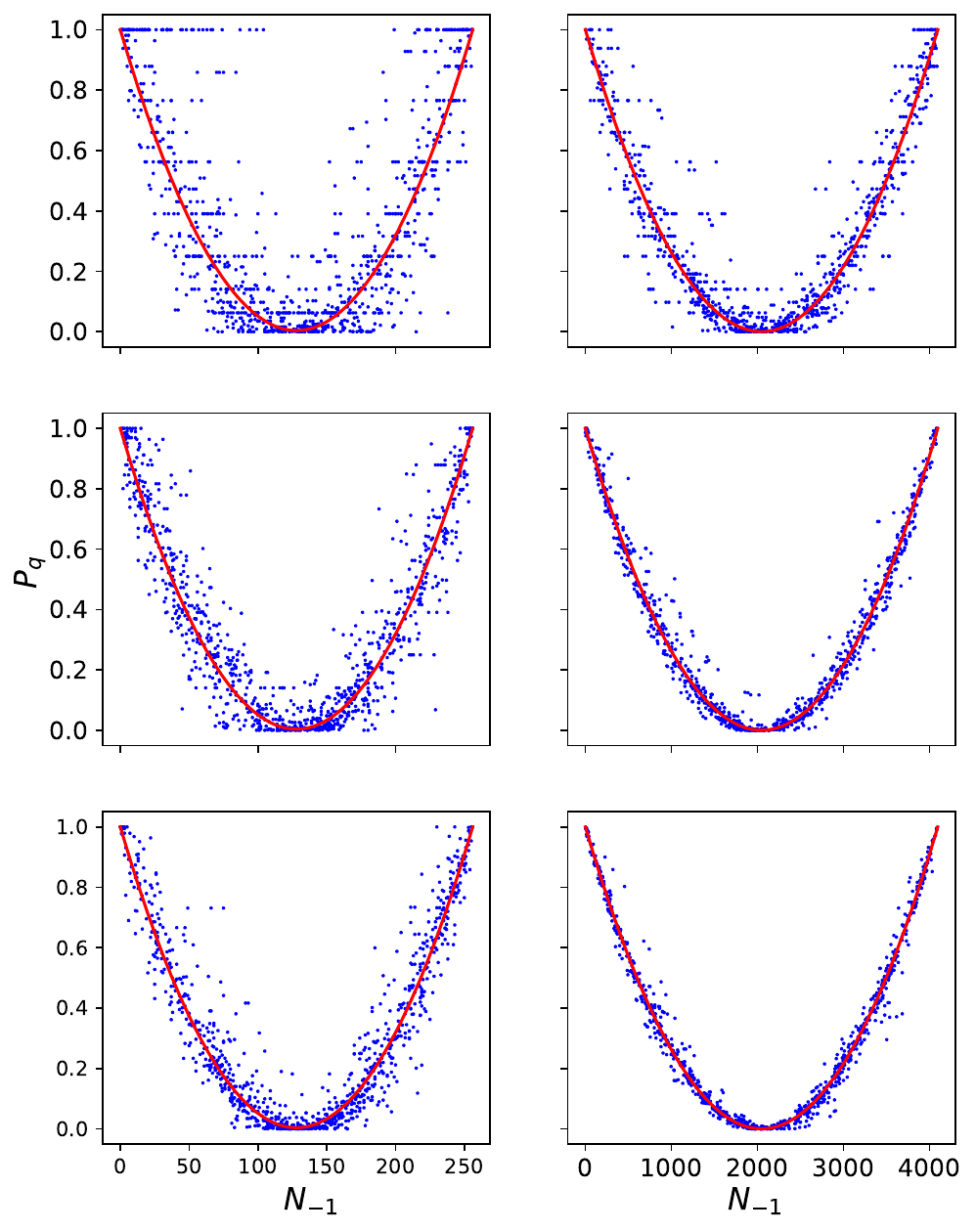}
    \caption{ \textcolor{black}{Probability, $P_q$ of measuring $\ket{00\cdots00}$ at the output of a randomly generated $n$-qubit zero discord unitary, with $\ket{00\cdots0}$ as the input state, plotted against $N_{-1}$. The solid line (red) is the analytical expression for the expectation value of $P_q$ from Eq.~\eqref{eq:var} while the dots (Blue) are the numerical obtained data points. For the plots on the left, $n=8$ and for those on the right, $n=12$. The unitary is of the form $V^{\dagger} D V$ where $V$ consists of $m$ layers, each constructed using sets of commuting gates randomly from the Clifford + $T$ gate set. The first, second and third rows correspond to $m = 4$, $m = 8$ and $m = 12$ layers respectively. We see that as $m$ increases the convergence of the values to the expected parabolic curve improves. Further, the convergence is better for larger $N$ for a given $m$. }}
    \label{fig:realcircuits}
\end{figure}
\textcolor{black}{In order to demonstrate that such a procedure can work, we start from a randomly chosen $D$ and see the behavior of $P_q$ when $V$ is built-up in stages using randomly generated circuit layers consisting of gates from a universal gate. We denote the number of layers in $V$ with $m$. We see that for small $m$, the unitary $V$ is more structured and the corresponding $P_q$ has a large scatter around the parabolic curve expected from Haar-random $V$. With the addition of more layers the probability converges to the expected parabola as observed in Fig.~\ref{fig:realcircuits}. Comparing similar computations for the 8-qubit and 12-qubit cases, we also observe that a higher number of qubits require fewer layers for better convergence. This behavior becomes useful in the next section where the BCH formula is applied to find an effective Hamiltonian. With increasing number of qubits, application of the BCH formula become computationally challenging but some of the difficulty can be offset if the number of layers required to construct $V$ is smaller. Note that, so far, in the numerical computations considered, the matrix $D$ is a $2^N \times2^N$ matrix with randomly placed $1$ and $-1$ values along the diagonal and need not always have an efficient implementation as a circuit consisting of one and two qubit gates (see Appendix \ref{appA}).}

\section{Classical Simulation}
We assume that the \textcolor{black}{input} unitary $U$ we are considering in the zero-discord case be \textcolor{black}{describable classically} as a quantum circuit with Poly($n$) elementary gates \textcolor{black}{and not as a matrix representation of $U$ \cite{2016Gu}}. If the DQC1 circuit itself is not efficiently implementable as a quantum computation then the question whether it can perform a task using exponentially faster than best known classical circuit is ill-defined. Resource counting in such scenarios must take into consideration the exponential cost in terms of time and energy required to realize all the gates required for implementing the circuit and this cost will grow exponentially if the circuit is not efficiently implementable. The celebrated Solovay-Kitaev theorem informs us that even if a specific unitary is not efficiently implementable, it can always be approximated to any desired accuracy with those that can be implemented efficiently~\cite{Kitaev_1997, Dawson2005} and so we restrict our attention to only such implementable unitaries. \textcolor{black}{This means that even if one happens to be interested in a DQC1 computation that involves a zero-discord unitary that is not efficiently implementable then one can approximate the computation using a suitable unitary that is implementable. Operationally this implies that restricting our attention only to implementable zero-discord unitaries in the following can be done without loss of generality.}

For implementable unitaries we assume that the quantum circuit using any preferred universal set of quantum gates~\cite{Barenco1995, BarencoJune1995} is available. Multiplying the gates together as matrices to find the unitary carries exponential computational cost and cannot be done. However our objective is served if a few of the diagonal entries of the unitary can be sampled efficiently. Using these values we can compute the sample mean $\langle U_{qq}^2\rangle = \langle O_{qq}^2\rangle \sim P_q$ and use it as an estimate of $P_q$. As can be seen from \ref{fig:P_state} if we know the value of $P_q$ which is on the $y$-axis, we can find $N_{-1}$ from the $x$-axis. The two possible values of $N_{-1}$ corresponding to each $P_q$ give the normalized trace of $U$ up to an overall factor of $-1$ which can further be disambiguated using the available circuit implementation of $U$. We outline below how a few diagonal elements of $U$ can be sampled efficiently using classical means. The method we propose does have one caveat though, which we will point out as we go along.  

The first step, if not already done, is to rearrange the gates of the given circuit implementation of $U$ into groups of mutually commuting gates acting simultaneously on the $n$-qubit register. For concreteness we further assume that $U$ is implemented using the universal gate set \textcolor{black}{(Clifford + T)} consisting of the three gates,
\begin{align*} 
{\mbox{Hd}}\!=\! e^{i \frac{\pi}{2} \big[ \openone - \frac{X+Z}{\sqrt{2}} \big]}\!, \,  T\!=\!e^{i\frac{\pi}{8} [\openone - Z]}\!, \, \\
{\mbox{\sc cnot}}\! =\! e^{i\frac{\pi}{4}[\openone - Z]\otimes[\openone - X]}, 
\end{align*}
where ${\rm Hd}$ denotes the Hadamard gate. Note that each of these gates can be written in the form $e^{ih_k^{(j)}}$ where $h_k^{(j)}$ are one- or two-local strings of Pauli operators multiplied by appropriate numerical coefficients. After the grouping the circuit has $m+1$ layers with each layer containing all the gates that can act simultaneously. Out of these one layer would correspond to the overall global phase with $e^{i\phi \openone}$ as the gate on every one of the qubits. We can read off the overall phase from this layer and focus on the remaining $m$ layers. The unitary transformation corresponding to the $j^{\rm th}$ layer of the circuit can be written as $O^{(j)} = e^{i\sum_k h_k^{(j)}} \equiv e^{iH^{(j)}}$, where $H^{(j)}$ are sums of products of Pauli operators with each $H^{(j)}$ having a Poly($n$) terms.

The DQC1 unitary, $O$ (modulo the global phase) can be written in terms of its circuit representation as
\begin{equation} 
\label{eq:UN}
    O  =  \prod_{j=0}^{m-1} O^{(m-j)} =   \prod_{j=0}^{m-1}  e^{iH^{(m-j)}}   \equiv  e^{iH},
\end{equation}
The operator $H$ is in general difficult to find and the problem of mapping a given quantum circuit to a local Hamiltonian is quite well studied~\cite{Kay2010HtoU,Nagaj2012}. However, here we have the additional advantage that $O$,  is an involution (Hermitian) with $O^2 = \openone$. The operator $H$ has to be constructed from $H^{(j)}$ through a judicious application of the Baker-Campbell-Hausdorff (BCH) formula \textcolor{black}{\cite{Achilles2012}}. Details of the convergence of the BCH expansion for the type of $U$ considered and its numerical exploration is included in Appendix~\ref{appA}. The operator $H$ is again a sum of Pauli strings.  We can now identify $O = V^{\dagger} D V = e^{iH}$ so that  
\begin{equation}
    \label{step1}
    D = V e^{iH} V^\dagger = e^{iVHV^\dagger},
\end{equation} 
is a diagonal matrix with $+1$ and $-1$ as entries. The operator $VHV^\dagger$ therefore can be considered as a diagonal matrix with $0$ and $\pi$ appearing as entries wherever $+1$ and $-1$ appears along the diagonal of $D$ respectively. Hence we have,
\begin{equation}
    \label{eq:specialform}
    D = -\frac{2}{\pi} VHV^\dagger + \openone^{\otimes n},
\end{equation}
which leads to,
\begin{equation}
    \label{eq:Oexpansion}
    O = V^\dagger D V =  -\frac{2}{\pi}H + \openone^{\otimes n}. 
\end{equation}

The matrix $O$ is essentially the sum of the the same set of Poly($n$) one- and two-local Pauli strings appearing in the Hamiltonian  scaled by $-2/\pi$. We now have to compute a fixed number of diagonal elements of $O$ to find the normalized trace of $U$ as discussed earlier. Since $H$ is sum of Pauli strings, each of which are also typically sparse matrices~\cite{Aharonov2003, lloyd}, this involves computing the diagonal elements at a fixed position of each these terms  essentially just by inspection of the Pauli strings that are involved. As the number of terms in $H$ are Poly($n$) albeit with a large multiplicative factor, computing a fixed number of elements $O_{qq}$ can  be done classically in an efficient manner~\cite{ Van2011SimulationPorb}, thereby completing our construction. \textcolor{black}{In Appendix~\ref{CL}, we have tested the algorithm and show that efficient classical computation of the trace of randomly generated zero-discord unitaries acting on $n$ qubits is possible for $n = 15$ and $n = 18$.} An important caveat in our discussion is that since $H\sim \log O$, $H$ is not, in general single valued. The key relationship in Eq.~(\ref{eq:Oexpansion}) holds only if $H$ belongs to the principal branch of the $\log O$ function. The construction of $H$ from the circuit decomposition of $O$ has to be such that at each step this condition is maintained. While it appears to be possible to do this in the case of \textcolor{black}{the types of zero-discord unitaries we have numerically explored with number of qubits ranging up to 20}, whether it can be done in general without having to evaluate $VHV^\dagger$ and looking at its $2^n$ diagonal elements remains an open question \textcolor{black}{beyond the scope of the present work}.  

\subsection{Schmidt Rank}
In~\cite{Animesh2007cor}, the Schmidt rank of the final state of the DQC1 circuit for a random implementable unitary was shown to scale exponentially as $2^{n/2}$ proving that classical simulation of the state using the approach in~\cite{vidal2003efficient} is not possible. For the implementable Hermitian unitaries we consider, the action of the unitary on a pure state $|\psi\rangle$ of Schmidt rank 1 is given by Eq.~\eqref{eq:Oexpansion} as essentially left multiplication by a \textcolor{black}{sum of Pauli operators with degree of locality bounded by the order of the BCH truncation}. The Lieb-Robinson bound~\cite{Lieb1972} tells us that each of the terms in the sum can increase the Schmidt rank of the state only by a fixed amount that is independent of $n$ and dependent only of the degree of locality of the operator. Since the number of terms in the sum is bounded by Poly($n$) we find that Schmidt rank of the output state of the zero discord DQC1 state is also bounded by Poly(n) indicating that methods from~\cite{vidal2003efficient} may also be adopted to simulate the DQC1 circuit in this case.

\section{Conclusions}
Equations \eqref{eq:means1} and \eqref{eq:Oexpansion} captures the key reasons that allow for an efficient classical simulation of the zero discord DQC1 circuit when $U$ is implementable. The first reason was that the magnitude of the normalized trace can be found from any diagonal element of $O$. The arguments that lead to Eq.~\eqref{eq:means1} can be generalized even when $U$ has more than two distinct eigenvalues and it can generate discord in the DQC1 circuit. However, we note that the convergence of $O_{qq}$ to ${\mathbb E}(O_{qq})$ is better if the number of distinct eigenvalues of $U$ are less in number. This is because even if we are nominally replacing $|v_{qk}|^2$ with ${\mathbb E}(|v_{qk}|^2) = 1/N$ to get Eq.~\eqref{eq:means1}, what we are actually doing is replacing $\sum_k|v_{qk}|^2$ with $N_{\pm}/N$ which is much better approximation when $N_{\pm} \gg 1$. This effect can be observed in Fig.~\ref{fig:P_state} where the agreement between the theoretical curve and numerically obtained data points for $N=4096$ is better in the middle ($N_\pm \sim 2048$) and at the ends where either $N_+$ or $N_-$ dominates. In the limit of the number of distinct eigenvalues of $U$ becoming equal to $N$, no such additional averaging will be possible and $O_{qq}$ can become a good estimate of its expectation value only by virtue of the law of large numbers which applies only for very large $N$. In comparison, the steps that lead to Eq.~\eqref{eq:specialform} and Eq.~\eqref{eq:Oexpansion} are not possible if $U$ has more than two distinct eigenvalues. This means that the techniques presented here cannot be extended to do efficient classical simulations of DQC1 circuit when $U$ generates discord between the fully mixed quantum register and the top qubit. 

In conclusion, we have proposed a classical algorithm using which we can efficiently estimate the trace of an implementable zero-discord unitary, in effect,  reproducing the statistics of a zero discord DQC1 circuit using Poly($n$) classical resources. We have shown that the conjecture made in Ref.~\cite{dakic2010necessary} that the zero discord DQC1 case may furnish an argument against considering quantum discord as a key resource behind mixed state quantum computing is not true when $U$ is implementable. There remains one caveat to the method we presented that needs to be fully addressed. Our results do not furnish a clinching argument in favor of considering quantum discord and other related measures which quantify non-classical correlations other than entanglement in mixed states as the key resource either but it considerably strengthens the argument and more importantly, restores quantum discord as one of the main contenders for such consideration. 
\begin{acknowledgments}
This work was supported by the the Department of Science and Technology, Government of India through a grant under the National Quantum Mission (Hub for Quantum Computation) and the Ministry of Education through the STARS grant MoE-STARS/STARS-2/2023-0809. The authors would also like to thank John George Francis for useful comments on the computational aspects of this work.
\end{acknowledgments}

\bibliographystyle{quantum}
\bibliography{bib}

\appendix

\section{ Computing \texorpdfstring{$H$}{H} \label{appA}}
 The results of our numerical exploration of repeated applications of the BCH formula for obtaining $H$ as a sum of a large, but Poly($n$) number of Pauli strings is summarized below. The first challenge is to identify or generate efficiently implementable unitaries with corresponding gate-based quantum circuits which have only $\pm 1$ as its eigenvalues. We outline a means of addressing this problem and following that the convergence of the BCH based approach and the number of Pauli strings in $H$ are discussed. 
 \subsection{Circuit Implementation for a Hermitian Unitary}
The task of generating random Hermitian Unitary Circuits is a non-trivial problem. To understand why, we look at the decomposition of a Hermitian Unitary provided in~\cite{Arabzadeh2016}. The Unitary is broken into three main parts by writing the unitary as $U=WDW^{\dagger}$. The central part of the circuit is the circuit decomposition of the diagonal $D$ and it is sandwiched by rotation gates and their corresponding conjugate gates on either side. The complexity of constructing a general diagonal Hermitian unitary scales exponentially in $n$ as shown in \cite{Houshmand2015, Zhang2023}, where the well-known decomposition of multi-control $C^k Z$ gates into $O(n^2)$ elementary gates \cite{Barenco1995} is used. But this method \cite{Barenco1995} relies heavily on performing arbitrary single-qubit rotations for the exact decomposition which is challenging to implement experimentally. It may be noted that fault tolerant implementation of single-qubit rotations are typically not possible as well~\cite{Welch2016}. To switch to a universal gate set such as the Clifford + $T$ gate set, requires applying the Solavey Kiteav Theorem and for multi-controlled $Z$ rotations, $C^k Z$,  exact decomposition is only known up to $k = 2$ \cite{Nielsen_Chuang_2010, decomp2024}. Due to this, we choose to construct our circuit for the diagonal part $D$ using randomly placed $Z$, $CZ$ and $CCZ$ gates only. On the left side of these gates we place $m$ layers of gates with each layer consisting of a collection of mutually commuting gates chosen from a universal gate set like Clifford + $T$. On the right side, there are again $m$ layers placed in reverse order with each layer consisting of the conjugates of the gates on the corresponding layer on the right side. Though this need not encompass the entire set of Hermitian Unitaries that are implementable efficiently on a quantum computer, we use this subset of Hermitian Unitaries to explore the convergence of repeated application of the BCH formula and the number of terms in $H$.
\subsection{Baker-Campbell-Hausdorff Convergence}\label{BCH}
We are interested in answering the question of how one arrives at the Hamiltonian, $H$ of the full Unitary efficiently, given the gate decomposition of the circuit. If that step can be achieved we can apply Eq.~(10) from the main text and calculate the normalized trace for a given circuit implementation of a Hermitian Unitary. To do this, we apply the BCH formulae to the set of random Hermitian circuits generated as described in the previous section and estimate how closely the Hamiltonian and the corresponding unitary approximation converges to the required unitary. The sandwich structure of the circuit, allows us to use the following form for the BCH formula with the pattern of coefficients appearing it leading to easier computations: 
\begin{align}
     e^{-sA} e^{B} e^{sA}  = & \exp \bigg( B + s[A,B] + \frac{s^2}{2!} [A, [A,B]]  \nonumber \\
      & \qquad \quad + \, \frac{s^3}{3!} [A, [A, [A,B]]] + \ldots \bigg)  
\end{align}

\begin{figure}[!htb]
    \centering
	\resizebox{4cm}{3cm}{\includegraphics{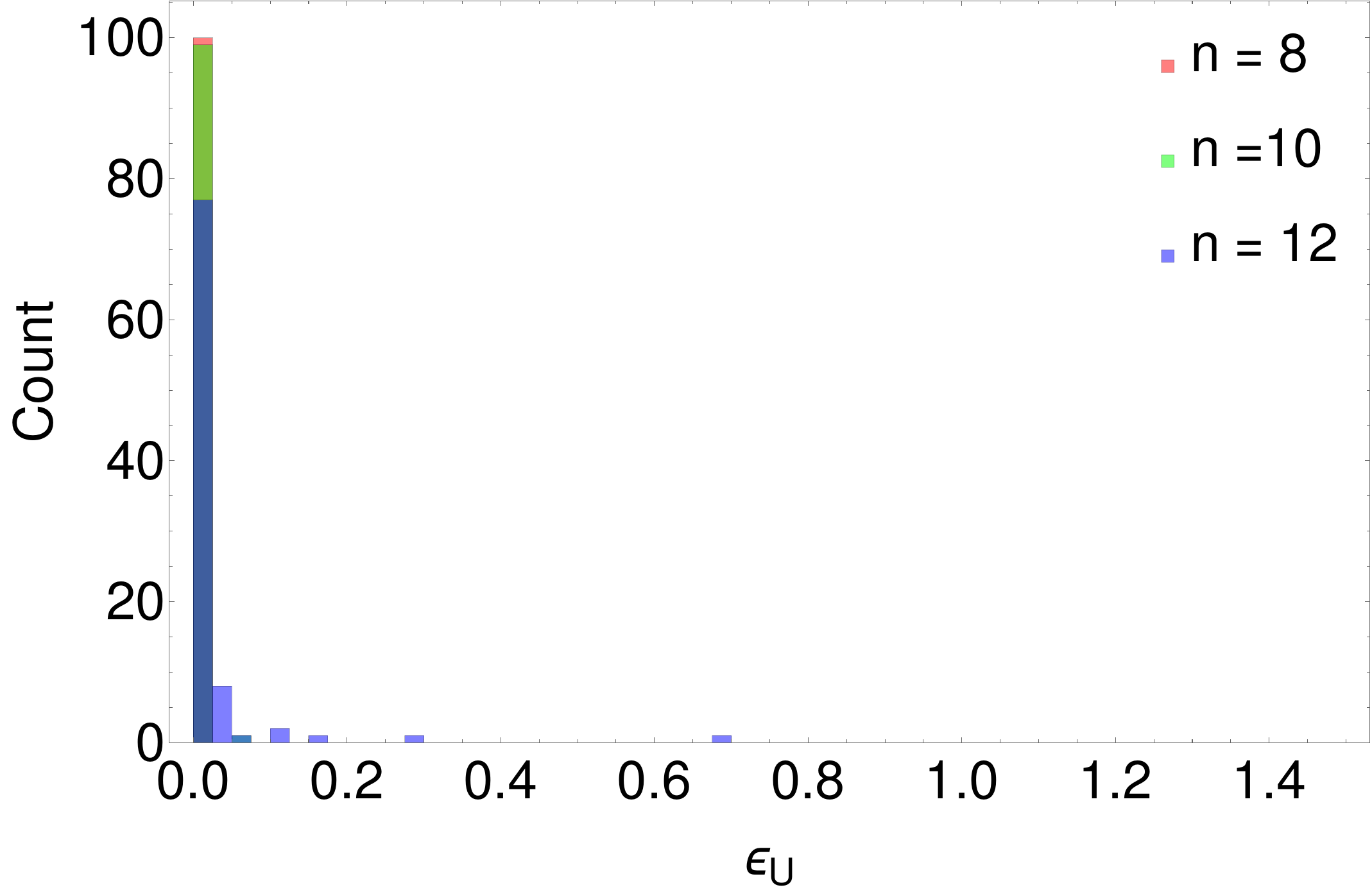}}
	\resizebox{4cm}{3cm}{\includegraphics{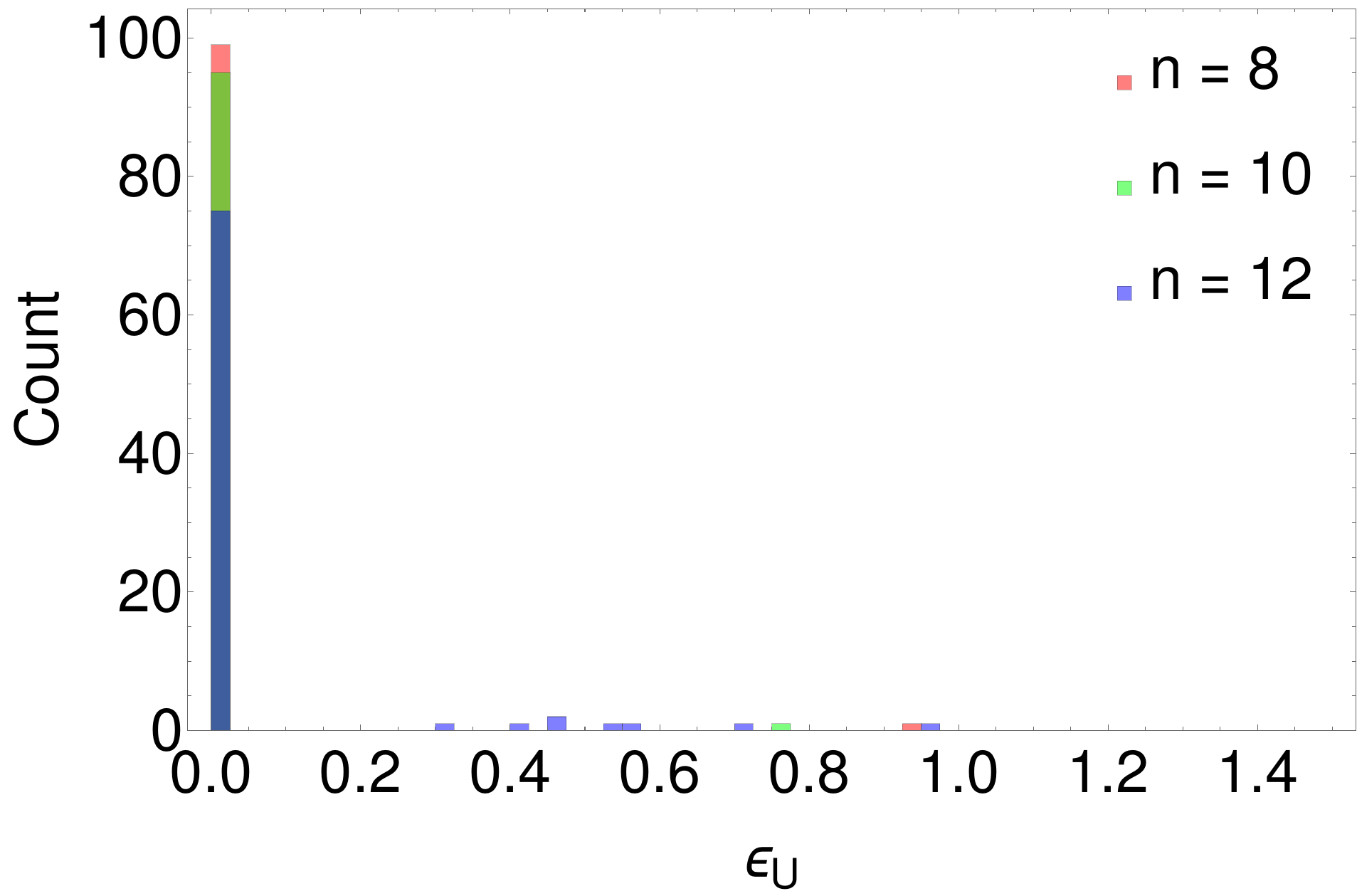}}
	\resizebox{4cm}{3cm}{\includegraphics{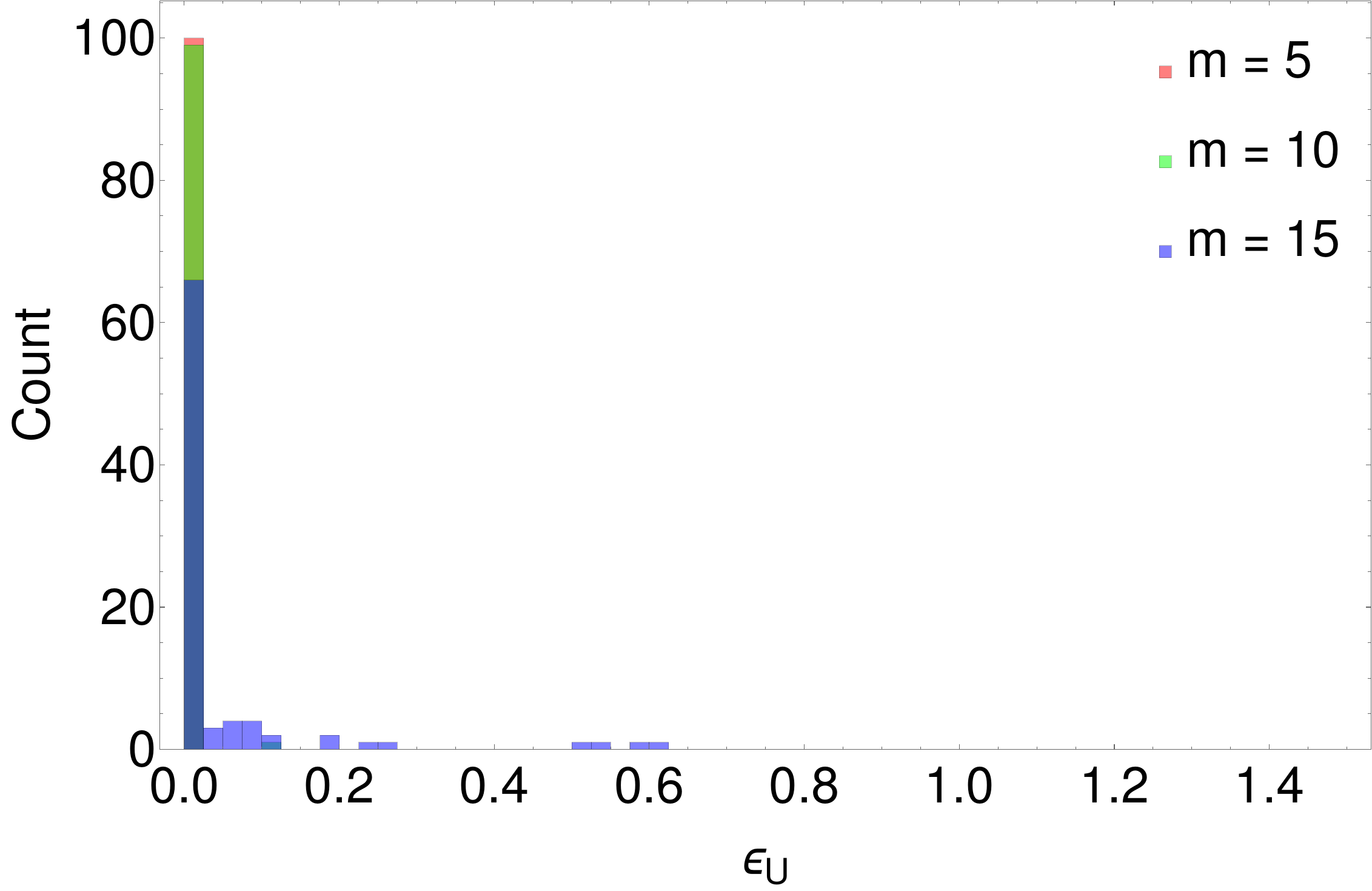}}
	\resizebox{4cm}{3cm}{\includegraphics{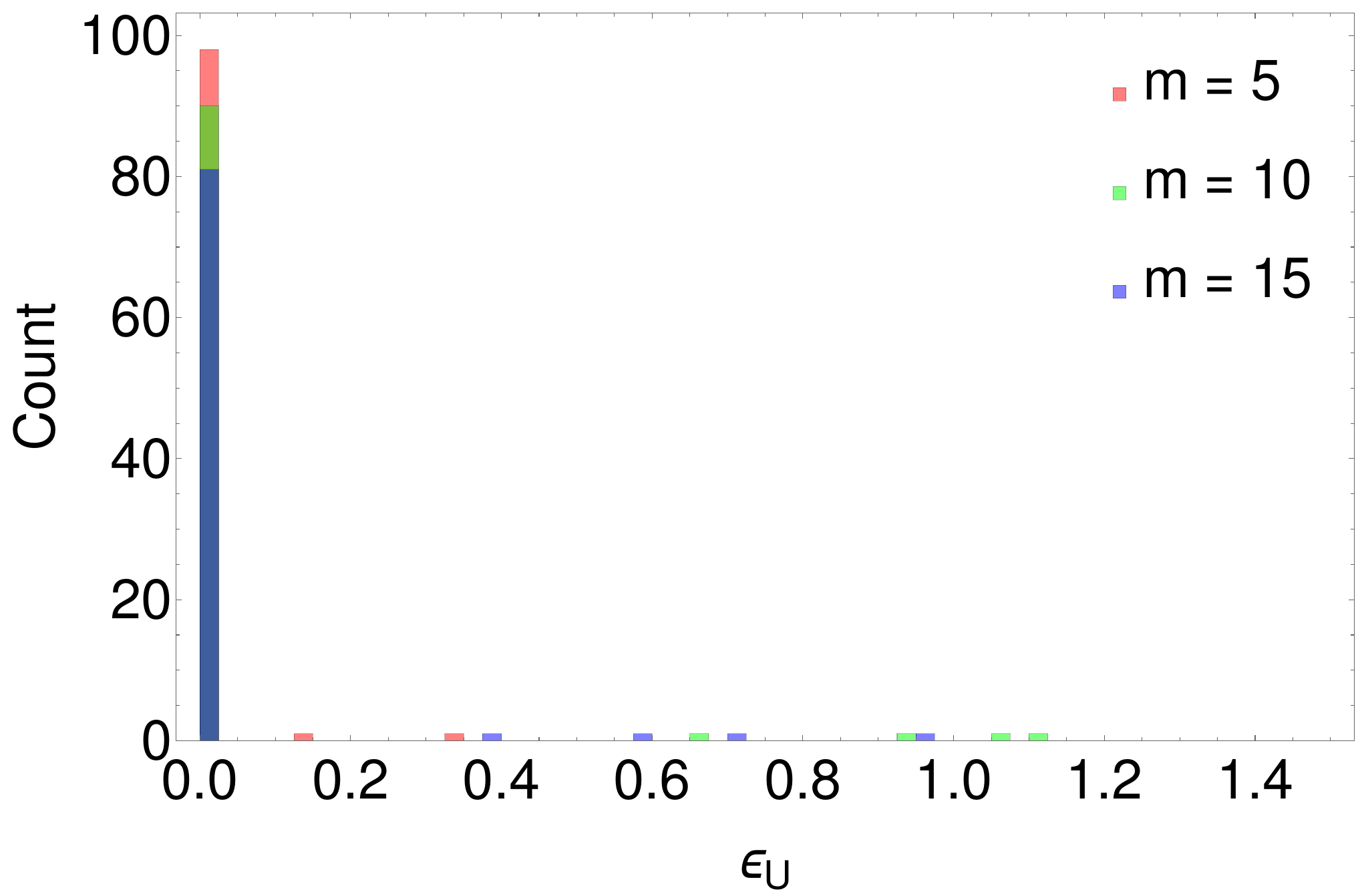}}
    \resizebox{4cm}{3cm}{\includegraphics{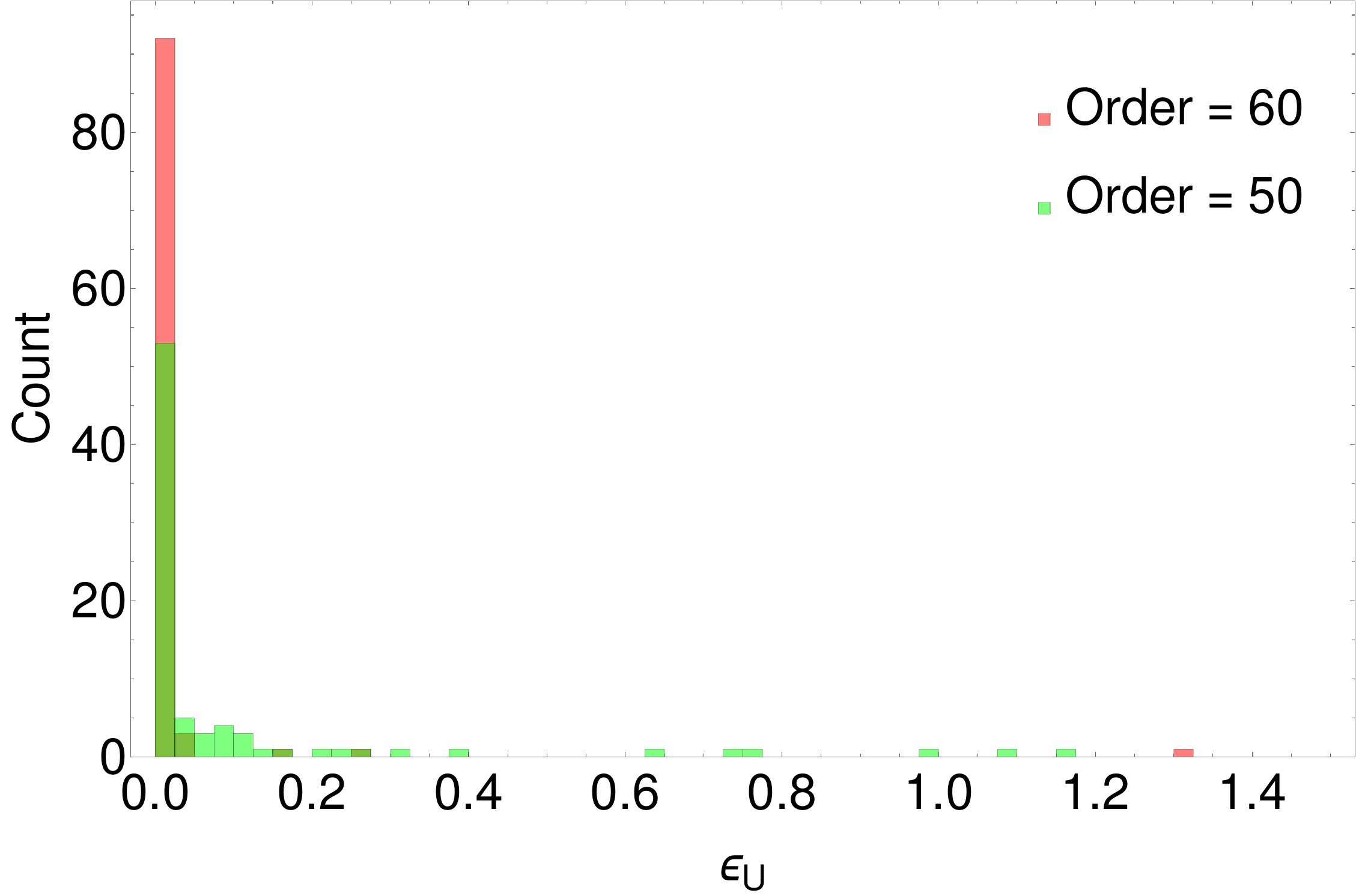}}
	\resizebox{4cm}{3cm}{\includegraphics{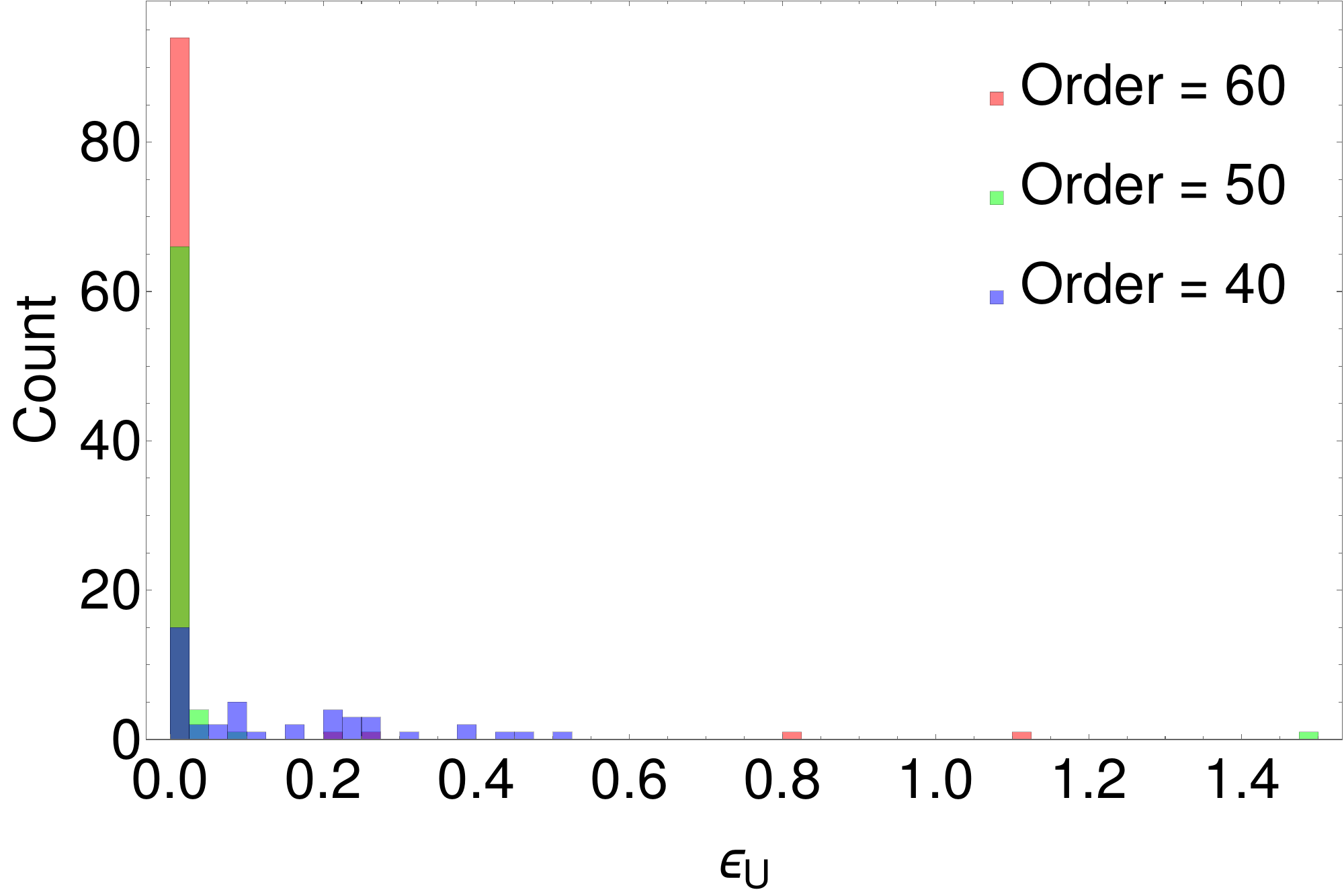}}
    \caption{The $L_2$ norm of the difference between the exact unitary and the unitary constructed from a Hamiltonian obtained using the BCH formula is plotted. For the plots on the left-hand side the expansion of the unitary was in terms of the gate set (CNOT + H + T) and on the right-hand side it was expanded using (Toffoli + H). The first, second and third rows shows convergence of the $L_2$ norm under BCH expansion for varying numbers of qubits, different number of layers in the circuit implementation of the unitary and the order of BCH truncation respectively. The sample size is 100 for each plot and the order of the BCH truncation for the first two rows is 60.}
\end{figure}  
The gates that are part of the diagonal matrix at the center all commute with one another and they are contained in the $e^B$ operator. The above expansion is repeatedly applied for each layer with truncation up to the specified order to estimate the final approximate Hamiltonian $H_{(1)}$ of the circuit. \textcolor{black}{We note that the eigenvalues of $H_{(1)}$ are equal to $0$ or $\pi$ only modulo $2\pi$. We discuss the special case where the eigenvalues are set to $0$ or $\pi$ separately.} The $ L_2$ norm of the difference between the exact $U$ and the approximate Unitary, given by $ \epsilon_u = |U - \exp{i H_{(1)}} |$, is plotted in Fig.(3) by varying different parameters such as $n$, $m$ and the order of truncation. This is done for gate sets, Clifford + $T$ and Toffoli + Hadamard. We find that the convergence to the expected Unitary varies only slightly with the gate set but they exhibit similar patterns. We also see that it converges around the truncation order 60 for $n = 8, 10$ for almost all the sampled circuits while for $n = 12$ it converges for most of the samples. We can expect it to converge for higher values of $n$ though we could only go up to 12 qubits as the computation gets heavy. The Unitary is also seen to converge faster for fewer number of layers as is expected. 

\subsection{Classical Simulation}\label{CL}
\begin{figure}[!tb]
    \centering
    \resizebox{7.8cm}{5.4cm}{\includegraphics{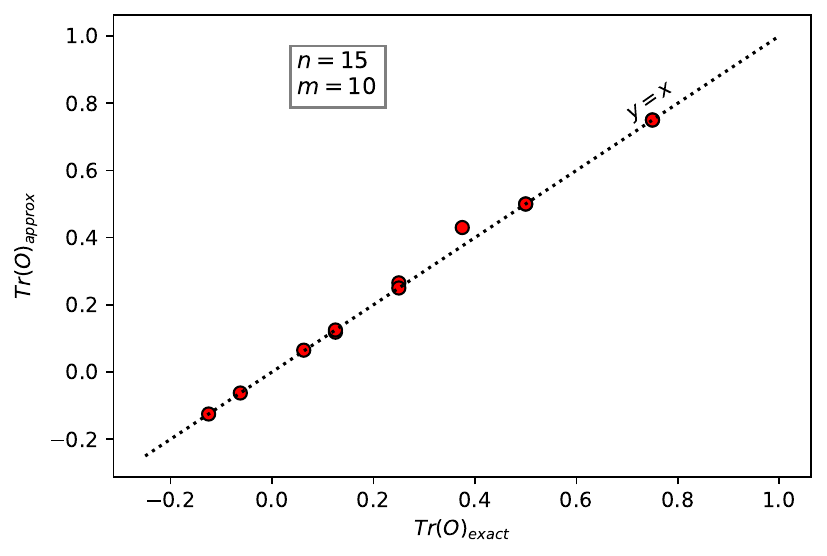}}
	\resizebox{7.8cm}{5.4cm}{\includegraphics{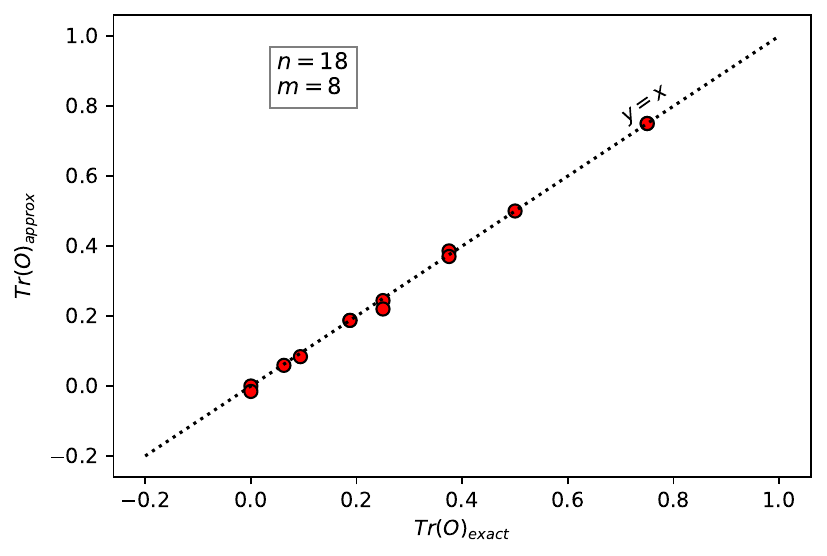}}
    \caption{\textcolor{black}{Normalized trace for randomly generated hermitian unitary circuits for $n$ = 15 and 18 with $m$ = 10 and 8 respectively is computed using the classical simulation algorithm and plotted (red dots). For each data point, 500 randomly selected diagonal elements of the unitary are average over. The data points have substantial overlap since many of the randomly generated unitaries have the same trace. The dotted line denotes the actual values of the traces and we see that the algorithm performs very well. }}
    \label{fig:fig4}
\end{figure}
The algorithm proposed was tested for a few randomly generated Hermitian unitary circuits using the method proposed in  Appendix~\ref{BCH}. For generating zero-discord unitaries required for verifying the algorithm, we have reverse-calculated the principle branch Hamiltonian using Eq.~(12) for a randomly selected sequence of diagonal gates that constitute $D$. The BCH expansion was then employed for each additional layer of gates (which are part of $V$ and $V^\dagger$ respectively) on either side.  Fig.~\ref{fig:fig4} shows a comparison between the exact and approximate trace values for $n= 15$ and $n = 18$ for $m = 10$ and $m = 8$ respectively. The averaging is performed on 500 randomly generated $D$ matrices. The computation was done in Julia using the PauliStrings.jl \cite{Loizeau2025}. Since the $D$ matrices were constructed using a limited set of gates, namely, $Z$, $CZ$ and $CCZ$, many of the matrices generated had the same trace. Similar calculations were done for $n =20$ whose results are not included, since we were able to perform only up to $m = 6$ in most cases. We observe that complexity of calculating the effective Hamiltonian via the BCH expansion increases with the number of layers $m$ on either side of the center block. This is because, although the number of Pauli terms in the effective Hamiltonian is Poly($n$), the polynomial could be of a high order depending on the layers being applied. 

\subsection{Scaling of number of terms of H}
\begin{figure}[!htb]
    \centering
    \includegraphics[width=0.9\linewidth]{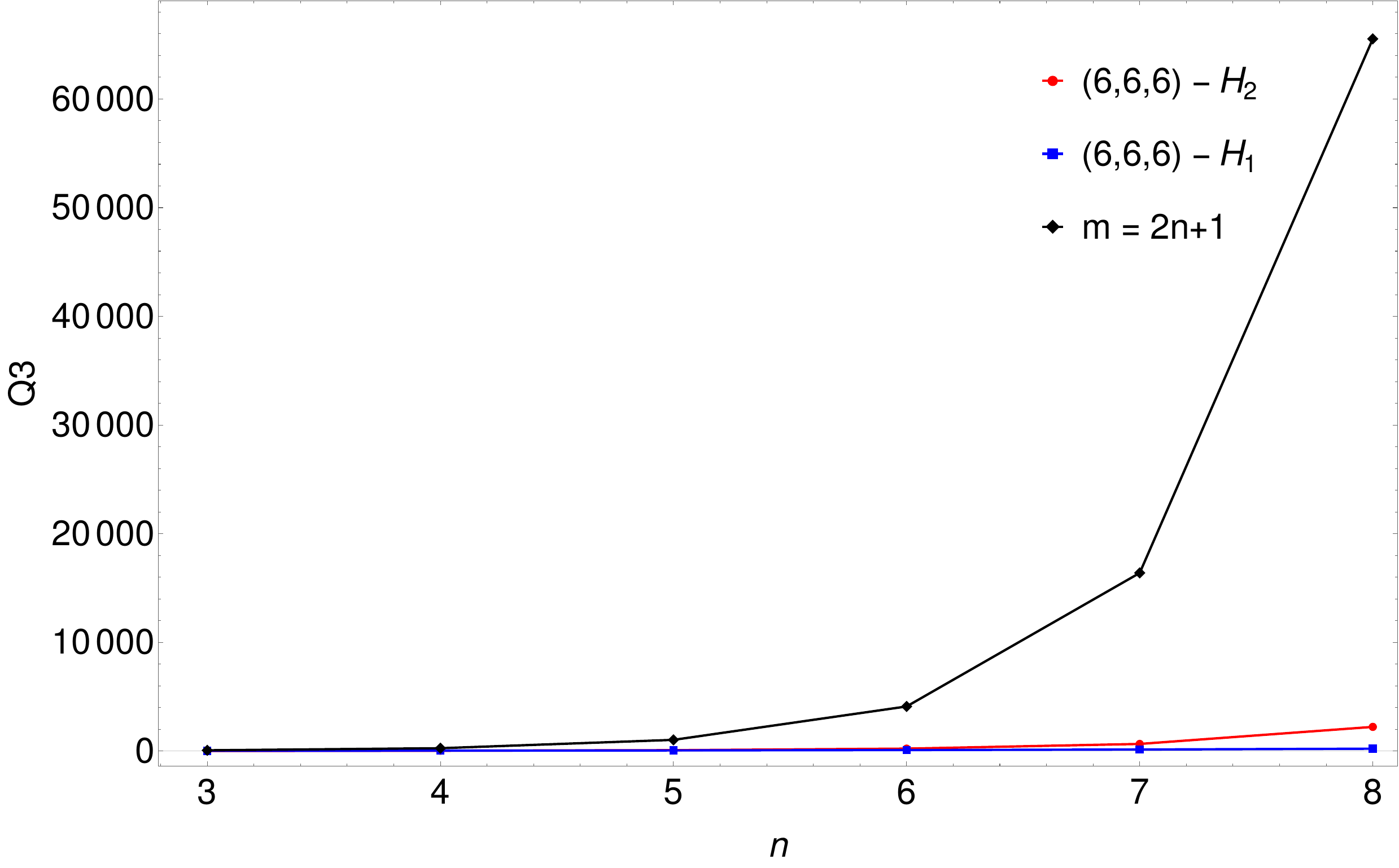}
    \caption{The third Quartrile ($Q3$) of the distribution of the number of Pauli strings in the Hamiltonian decomposition as a function of the number of qubits $n$. $H_{1}$ and $H_{2}$ are blue and red lines plotted for $(n_{ccz},n_{cz},n_z) = (6,6,6)$ respectively while the black line indicates the case of random circuit Hamiltonians. The sample size is 1000 for each $n$.}
    \label{fig:fig5}
\end{figure}
\textcolor{black}{The classical trace estimation method outlined here depends on the number of Pauli strings that make up the Hamiltonian operator and it would fail if the number scales exponentially in $n$. Therefore, we obtain numerical estimates for the number of terms in the Hamiltonian operators $H_{(1)}$ and $H_{(2)}$ corresponding to the same Hermitian unitary given in Fig.~\ref{fig:fig4}. Here, $H_{(2)}$ is the Hamiltonian that satisfies Eq.~(9) of the main text and has eigenvalues 0 or $\pi$ only. Therefore, $H_{(2)}$ also reflects the growth of the number of terms in $O$ as given in Eq.~(10) of the main text}. A comparison is shown between Hermitian circuits and randomly generated circuits in Fig.~\ref{fig:fig5}. The Hermitian circuit is constructed as described above. For random circuits, we have fixed $m = 2n + 1$ layers and decomposed its matrix logarithm into a Pauli string summation. We have sampled 1000 circuits for each $n$. Although we cannot exactly determine the order of the growth rate of the number of terms in $H_{(2)}$ for Hermitian circuits, it is much slower compared to the number of terms in the randomly generated circuits which grow as $O(4^n)$. $H_{(1)}$ is also plotted for comparison, as it is known to scale as a polynomial in $n$ by the nature of its construction. This would suggest that the Hamiltonian description of the Hermitian unitary operators and therefore the output state of such unitaries can be computed classically using far fewer resources with increasing number of qubits compared to a randomly generated unitary and its corresponding Hamiltonian.

\end{document}